# Earth as a Hybrid Planet: The Anthropocene in an Evolutionary Astrobiological Context

Adam Frank[1], Marina Alberti[2], Axel Kleidon[3]

[1]Department of Physics and Astronomy, University of Rochester, Rochester, New York, 14620
[2]Department of Urban Design and Planning, University of Washington, Seattle, Washington, 98195
[3]Max Planck Institute for Biogeochemistry, Jena, Germany

**Abstract**

We develop a classification scheme for the evolutionary state of planets based on the non-equilibrium thermodynamics of their coupled systems, including the presence of a biosphere and the possibility of what we call an "agency-dominated biosphere" (i.e an energy-intensive technological species). The premise is that Earth's entry into the "Anthropocene" represents what might be, from an astrobiological perspective, a predictable planetary transition. We explore this problem from the perspective of the solar system and exoplanet studies. Our classification discriminates planets by the forms of free energy generation driven from stellar forcing. We then explore how timescales for global evolutionary processes on Earth might be synchronized with ecological transformations driven by increases in energy harvesting and its consequences (which might have reached a turning point with global urbanization). Finally, we describe quantitatively the classification scheme based on the maintenance of chemical disequilibrium in the past and current Earth systems and on other worlds in the solar system. In this perspective, the beginning of the Anthropocene can be seen as the onset of the *hybridization of the planet* - a transitional stage from one class of planetary systems interaction to another. For Earth, this stage occurs as the effects of human civilization yield not just new evolutionary pressures, but new *selected* directions for novel planetary ecosystem functions and their capacity to generate disequilibrium and enhance planetary dissipation.

I. Introduction

The recognition that human activities alter Earth's climate has prompted debate concerning "planetary boundaries" (Rockstrom 2009, Barnosky 2012, Lenton 2008, Brook et al 2013) required to keep the anthropogenic forcing (Steffen et al. 2015) within "safe operating limits". These studies concern Earth's entry into a possible new geologic epoch called the "Anthropocene," where humanity's collective actions become the dominant driver for planetary changes (Crutzen 2002). Here we refer to the "Anthropocene" as a formal unit in the geological timescale defined by the 'Anthropocene' Working Group and under consideration by the International Commission on Stratigraphy (Zalasiewicz 2017).

While some researchers question if the Anthropocene can be defined via stratigraphy (Zalasiewicz 2015), substantial evidence exists that Earth has already crossed a boundary where key measures show large-scale human signatures. As two examples, we note that more than 50% of the Earth's land surface area has been "colonized" for human uses (Hooke et al 2012), and current anthropogenic flows of phosphorus are more than factor of 5 above "natural" rates (8 Tg P $y^{-1}$ anthropogenic vs 1.1 Tg P $y^{-1}$ natural; 5). Thus more generally one can define the Anthropocene as a new epoch in which human effects dominate many of the coupled Earth Systems. We will use the term in this sense in what follows.

Earth's entry into an anthropogenic era poses challenging questions for the long-term sustainability of global human civilization. It is, in fact, not clear if a planetary civilization as energy-intensive as ours can be sustained for centuries. While some aspects of this question rest within political science and sociology (Kennett & Beach 2013), a broader perspective is developing on the transition, that only a new collaboration among the physical, biological, and social sciences can address to illuminate and inform the choices we face. In what follows we will consider a perspective which can be called the *Astrobiology of the Anthropocene*. This entails viewing the transition the Earth and the human project of civilization are currently undergoing in their full astronomical and planetary contexts. This means seeing our project of civilization as another manifestation of the long co-evolution of the biosphere and other coupled Earth systems. It also means broadening that view to ask if what we call the Anthropocene might be a generic consequence of any planet evolving a successful technological species. We believe this perspective holds considerable benefits in understanding the true nature of the challenges we face and articulating paths towards solutions. We will develop this perspective and articulate its benefits further in what follows.

Understanding of the current state of the coupled Earth systems has not developed in isolation. Only through intense study of Earth's 3.8 billion year history of habitation have we gained insights into the strong and complex interplay between the biosphere and other systems (Kasting & Canfield 2012, Arndt & Nisbet 2012). The study of other solar system bodies has also provided powerful laboratories for understanding climate, in terms of radiative transfer, atmospheric chemistry, atmospheric dynamics and couplings to geological and near-planet space environments (Rockstrom 2009). The recent explosion in exoplanet studies has also become relevant. Researchers are on the threshold of exoplanet atmospheric characterization, in which information concerning the chemistry, dynamics and, perhaps, the presence of bio-signatures, for these worlds is expected to be forthcoming (Howard 2013, Lineweaver & Chopra 2012, Seager, S. 2013).

In light of these advances, it is now possible to cast the question of boundaries and thresholds for Earth systems into a wider context concerning life and its planetary environment. Instead of focusing purely on human impacts on Earth, it should now be possible to develop implicitly astrobiological frameworks for broadening our understanding of coupled system dynamics on *any* planet with *any* level of biosphere (Frank & Sullivan 2014). From this perspective, one can develop a coherent account of the "rules of the game" for different planetary system interactions that are quantifiable and testable. Developing even the outlines of such rules would be of scientific interest in their own right. In addition, understanding general features of such dynamics can also help define boundaries and thresholds we may be facing with our own impacts on Earth.

Implicit in this framework is an assumption that Earth would not be the only world on which life evolves. Indeed, our world would not be the only one to evolve an energy-intensive technological species with the capacity for planetary-scale feedbacks.  Clearly, the existence of both life and intelligence on other worlds represents one of the greatest open questions in science.  The perspective we take here, however, assumes a fairly conservative answer to these questions (Frank & Sullivan 2014, 2016).  Such evolution only needs to occur a statistically relevant number of times (~ 1000) across all cosmic time and length scales for *average properties* of the kind we are interested in here to be relevant. As long as the probability for a habitable planet to develop a technological species once in its history is greater than $10^{-19}$, then meaningful averages must exist (Frank & Sullivan 2016).

This paper takes a global, systemic, and inherently astrobiological view of sustainability for any energy-intensive technological civilization.  Viewing planets as thermodynamic systems, we first develop a classification scheme for different levels of coupled planetary systems in terms of their rates of free energy generation.  Next, we build on current evidence of ecosystem-evolutionary dynamics to explore how biospheric-planetary systems interactions can evolve novel ecosystem functions, and apply this to "anthropocene"-like transitions. We conclude with a description of how the proposed framework can apply in future investigations.

## II. Purposes of Planetary Classification: Beyond the Kardashev Scale

We address the purpose in developing a new classification scheme for planets with life with an emphasis on energy-intensive civilization building species.  Many classification schemes have been developed in the history of astrobiology, and the Search for Extraterrestrial Intelligence (SETI in particular, with the Kardashev scale  particularly influential (Cirkovic 2015).  At first glance, our approach bears similarity to the Kardashev scale.  Our strategy is substantively different, however, in both its intent and its understanding of the evolution of technological civilizations. It explicitly links to energy conversions within the planetary environment and how these are constrained by thermodynamics.  These differences are important for filling gaps in previous research.

The Kardashev scale was first proposed (Kardashev 1964) as a means of classifying technological civilizations ( we refer to them below as *exo-civilizations*). The purpose of the Kardashev scale was to aid discussions of detectability through SETI efforts. The scale was based on the exo-civilization's energy consumption/manipulation levels. A Type 1 civilization manipulated the entire energy resources of its home planet. A Type 2 civilization manipulated the entire energy resources of its home star/planetary system. A Type 3 civilization manipulated the entire energy resources of its home galaxy. The Kardashev scale was originally intended to help guide SETI by classifying its range of possible targets on an evolutionary scale. Since then, however, it has also become a kind of gold standard for thinking that focused purely on the evolution of exo-civilizations. The literature on the Kardashev scale is long, encompassing topics from enhancements (Galantai 2004), criticisms (Galantai 2007), engineering (Armstrong and Sandberg 2013) and philosophy (**Barrow 1999**).

Our proposed classification system is not intended as a means for structuring exo-planet observational programs (though some may find it useful for such an endeavor). Instead, this scheme will structure a research program for understanding the trajectories of co-evolution of planets and life, explicitly to include development of what we call *agency-dominated* biospheres, i.e. a sustainable exo-civilization. Our thesis is that the development of long-term sustainable versions of an energy-intensive civilization must occur on a continuum of interactions between life and its host planet. Developing this classification system would lay the foundations for future work on the co-evolution of life and planets along this continuum. Thus, our research framework takes an explicit perspective in which long-term sustainable civilizations are not seen as "rising above" the biosphere. Instead, the path to long-term sustainability demands learning how to "think like a planet" (Alberti 2016), by entering into a co-operative ecological-evolutionary dynamic with the coupled planetary systems.

In this way, our classification differs from the Kardashev scale and its literature. The Kardashev scale originated from a particular historical moment in thinking about exo-civilizations, in which technology would be unconstrained, hence its focus on energy consumption alone. Civilizations were expected to rise up the ladder of energy manipulation while the physical systems from which that energy was drawn would simply be brought to heel. In this way, considerations of Type I civilizations represent a kind of planetary brutalism complete with implicit visions of world-girdling cities (e.g. Trantor in The Foundation trilogy (Asimov 1951).

In the years since Kardashev proposed the classification system, we have learned (the hard way perhaps) that biospheres are not so easily ignored. From the work of Lovelock, Margulis and others (Lovelock 1975, Lovelock & Margulis 1974) , a new scientific understanding of planets and life has emerged that includes recognition of their co-evolution as coupled complex systems. Those systems have their own internal dynamics which must be considered when mapping out trajectories of civilizations as a form of biospheric activity arising in a planet's evolution. Thus, it is not simply energy consumption which must be considered. A thermodynamic perspective which includes the fundamental limits to how energy can be generated as well as the consequences of using that energy, i.e. entropy and free energy gradients, must be included in order to understand how civilizations rise to the level of Type I and, possibly survive long enough to move beyond their host world towards a Type II.

Thus, our proposed classification implicitly includes another stalward of the astrobiology literature: the Drake Equation (Vokoch & Dowd 2015). In particular, it is the final factor in that equation, the average lifetime of exo-civilizations ($\bar L$) which is at issue. It is not yet clear that *any* long term sustainable version of our kind of civilization (rated as 0.8 on the Kardashev scale) is even possible. This would imply a low values the average lifetime: L ~ 100 - 1000 y. This is clearly a question relevant to the Fermi Paradox and so-called "Great Silence". A vast literature exists on these issues (Brin 1983, Cirkovic 2009), including the role of catastrophes and the Kardashev scale. These studies do not, however, address the thermodynamic and eco-evolutionary issues raised with our classification scheme. In particular, they do not lay out a framework for putting civilizations back within the proper context of the evolution of planetary biospheres.

To summarize, our classification scheme is not intended to provide a new framework for exo-planet observations. It is not meant to be a means to find exo-civilizations. Instead, it provides a framework for understanding the "Anthropocene" we are now experiencing on Earth in a general context, By this we mean understanding the strong forcing of the planetary systems by human civilization as a potentially generic phenomena that will occur for any planet evolving an energy intensive technological species. The goal of this work is also to offer a program that might aid in understanding what sustainable outcomes must look like. In other words, if one does not know where one is going, it will be hard to get there.

We now provide the theoretical basis for our classification system and enumerate those classes.

### III. The Five Classes of Planets

We develop a classification based on the magnitude by which different planetary processes – abiotic, biotic, and technologic – generate free energy, i.e. energy that can perform work within the system. Most importantly, these different forms of free energy reflect states of thermodynamic disequilibrium. Examples of such disequilibrium states are (Kleidon, A. 2016): kinetic energy associated with atmospheric motion; unsaturated air over a water surface; the chemical composition of the atmosphere with its high abundance of oxygen and organic biomass at the surface.

Using disequilibrium as a metric for a planetary systems evolutionary state, we then use Earth's 4.5 Gy history, as well as other solar system bodies (planets and moons), to outline four planetary system classes. In addition, current levels of human activity formulated in terms of free energy also allow us to anticipate what might constitute a 5th planetary class where the activity of an energy-intensive technological species drives planetary systems in a sustainable manner.

*III.1 Thermodynamic Background*

Our classification utilizes a combination of thermodynamics, specific planetary conditions, and linkages between planetary sub-systems (Kleidon 2010, 2012, 2016). To understand how planetary systems perform physical (or chemical) work and generate disequilibrium states, we consider processes generating physical, chemical, and biologically related forms of free energy from stellar radiative forcing (Fig. 1). In principle, energy could also be generated by nuclear fission and, possibly, fusion, which could provide an additional source of free energy. We excluded it from consideration here, however, and focus on sustainable forms of free energy generation. We also consider how energy-intensive technological species can drive free-energy generation, e.g. with technology such as photovoltaic cells.

The flux of incident stellar radiation constitutes the principle thermodynamic forcing for planets. The first law of thermodynamics states that this energy is being conserved when converted. Energy conservation does not, however, indicate what type of energy conversions and associated dynamics takes place *within* coupled planetary systems. These internal planetary dynamics result from and are constrained by the second law of thermodynamics. In a non-isolated system such as a planetary system that exchanges radiation, the second law must be evaluated in the context of the planetary entropy balance. This balance relates the change in the entropy of the planet ($S_p$), with entropy exchange by radiation as well as entropy production that takes place within the system:

$$\frac{dS_p}{dt} = \frac{J}{T} + \frac{J_c}{T_c} + \sum_i \quad \sigma_i => 0 \qquad (1)$$

where $J$ = stellar radiative energy flux; $J_c$ = planetary radiative flux, $T_h$ and $T_c$ the temperatures at which the stellar and planetary radiative fluxes are emitted. Thus, the first two terms on the right side are fluxes of radiative entropy associated with stellar and terrestrial radiation. The third term is the one that matters for our consideration, as it represents the sum of entropy production by internal dissipative processes. In the steady state entropy balance, this term represents the entropy production that results from all forms of planetary dynamics: frictional dissipation of motion; chemical reactions; biotic respiration; energy use by a technological civilization. In this way, internal entropy production acts as a constraint on how much work can be performed within the system.

Planetary entropy production, however, does not discriminate processes that do not involve work (absorption of radiation and its re-emission or heat diffusion) from processes involving the generation and dissipation of free energy, as is for instance the case with motion that involves the generation of kinetic energy and its frictional dissipation.  This focus on free energy generation is the basis for our thermodynamic classification of planetary environments.  Planetary energy and entropy fluxes from stellar radiation set a global potential for the generation of free energy, denoted by $P_{rad}$. To evaluate how much of this potential can be used to generate forms of free energy, we need to consider the linkages of the planetary environment as shown in Fig. 1.

The starting point for such free energy generations is the connection between radiative fluxes and motion within the coupled planetary systems.  On planets with greenhouse atmospheres, the radiative heating and cooling take place at different places and times. Radiative heating of the surface and cooling of the atmosphere aloft sets up temperature differences driving atmospheric convection.  Convection is associated with the conversion of a fraction of the differential radiative heating into planetary motion, equivalently to the work heat engines perform.  This work generates free energy, (i.e kinetic energy), and a physical form of disequilibrium via velocity differences.  We denote this free energy by $P_{clim}$, as most of it is associated with the climate system dynamics.  Note that the rate by which work can be performed to generate planetary motion depends on the specifics of a planet.For instance, it depends on whether the planet is tidally locked, the extent to which it is tilted, and whether it has water that can evaporate and that can drive moist convection upon condensation and change radiative exchange by formation of clouds.  These factors affect how differential radiative heating takes place in a specific environment,  precluding simple, general expressions.

When motion is generated in a planetary environment, it yields transport, mass mixing and enhanced chemical activity.  These processes drive hydrologic cycling associated with further conversions of energy. Such conversions include generation of chemical free energy such as desalinization of rainwater, lightning-based nitrogen fixation, or the weathering of surface rocks.  Other means of chemical free energy generation include photochemistry, such as stratospheric ozone production via ultraviolet absorption and geochemical cycling due to the dynamics of planetary interiors. Together, these mechanisms generate chemical free energy and disequilibrium whose total generation rate we call $P_{chem}$.  The particular rates depend on the specifics on the planetary environment, e.g., the atmospheric composition and the presence of an oxic atmosphere, the presence of water, and a planet with plate tectonics as these affect the different mechanisms by which chemical free energy can be generated in the planetary environment.

The presence of photosynthetic life directly converts a fraction of stellar radiation into chemical free energy.  We denote this generation rate by $P_{bio}$.  Note that the need for nutrients in producing biomass links biotic activity to $P_{clim}$ and $P_{chem}$ since they provide and mix chemical constituents.

Last, but not least, the activity of a technological civilization relates to free energy generation processes by its energetic requirements for metabolic and socioeconomic activity.  Human appropriation of net primary productivity (Laland 2014) consumes some of the free energy generated by the biosphere to meet food demands while the consumption of fossil fuels drives socioeconomic activities.  Renewable energy draws from these forms of energy that are generated within the planetary environment.  For instance, solar power draws directly from the potential $P_{rad}$, while wind energy utilizes a fraction of $P_{clim}$.  We denote the energy appropriated for use by a technological civilization as $P_{civ}$.

### III.2 Class Specification

We  classify planets based on the magnitude of free energy generation by the different processes from the radiative forcing to the energy appropriation of a technological civilization.  The presence or absence of these generation processes, as well as the strength of their coupling to the planetary forcing,  naturally separates planets into different classes (Fig 2, 3).

**Class I:** Planets without an atmosphere are characterized by states close to radiative equilibrium. For these worlds, a simple energy balance determines the surface temperature, in which absorbed stellar radiation approximately equals the emitted radiation from the planet (Fig 2). Rotation, or the lack thereof, as well as heat storage provides an additional complication. Without an atmosphere with greenhouse gases, however, no mechanisms exist by which energy and entropy can drive significant energy conversions associated with additional planetary processes. Thus, physical power is absent ($P_{clim}$ = 0), as is chemical and biotic power ($P_{chem}$ = $P_{bio}$ = 0). The entropy production of the planetary system entirely results from radiative processes, in particular absorption and emission. Examples of Class I planets in our solar system are Mercury and the Earth's Moon.

**Class II:** For planets hosting atmospheres containing greenhouse gases, incident solar radiation creates thermal gradients between the surface and atmosphere where radiation is re-radiated back into space (Fig 2b). These gradients drive the generation of kinetic energy associated with convective atmospheric flows. Differential stellar energy deposition will also drive latitudinal and longitudinal circulation patterns. Particulate transport and chemical imbalances then follow leading to long-term disequilibrium. Thus, Class II planets show significant power in geo-climate systems ($P_{clim}$ >> 0), possibly some chemical power ($P_{chem}$ ≥ 0), but no biotic activity ($P_{bio}$ = 0). The entropy production of the planetary system in this class contains contributions by frictional dissipation, chemical reactions and other planetary processes. Venus and Mars in their present states represent Class II planets.

**Class III:** Biotic activity on a planet can be sustained by the use of chemical free energy from the environment, or by generating free energy out of stellar radiation. It is possible that a planet may host what we term a "thin" biosphere meaning while it sustains biotic activity, it does not strongly affect planetary drivers and alter the evolutionary state of the planet as whole (Fig 2c). Thus Class III planets have $P_{clim}$ >> 0, $P_{chem}$ > 0, and $P_{bio}$ >

0 but P$_{bio}$ does not result in strong feedbacks to the planetary forcing.  There are no current examples of Class III planets in our solar system, but Earth in the early Archean after life formed but before the great oxidation event may have represented such a world.  Should Mars have developed life during the Noachian, when liquid water existed on its surface, then it too may have represented a Class III planet.

**Class IV:**  Once life strongly feeds back to the radiative forcing of a planet, it becomes a Class IV world.  It hosts what we call a "thick biosphere" meaning all systems are strongly modified by life and that continual modification drives processes maintaining planetary disequilibrium (Fig 2d).  Earth's "thick biosphere" is sustained by photosynthetic activity as it requires sunlight as an energy source.  The presence of life, as Lovelock, Margulis (Lovelock & Margulis 1974) and others (Kliedon 2010, 2012, 2014) argued, dominates the coupling between planetary systems and results in planetary change, e.g., by altering the rate of chemical weathering (Schwartzman & Volk 1989) or surface energy and water balances on land (Shukla & Mintz 1982, Kleidon et al 2000).  Thus, Class IV planets are characterized by a greater share of biotically-generated power (P$_{bio}$ >> 0) and strong feedbacks with the planetary radiative forcing.  Earth after the great oxidation event represents a Class IV planet.

**Class V:**  As the final class, we imagine a planet in which the activity of an energy-intensive technological species strongly shape free energy generation and feedbacks . Such a class is possible because the radiative planetary forcing can, in principle, be converted directly into free energy with a huge potential ($P_{rad}$), but only if the intermediate step of radiative heating (i.e. the conversion of radiation into heat, e.g. by absorption at the surface) is prevented from taking place.  Photovoltaics  can accomplish this step as a technological means to generate free energy from solar radiation unavailable to natural processes.  Such a state would be characterized by substantial generation of free energy by technology so that $P_{civ}$ >> 0, and would be strong enough to affect the planetary radiative forcing.  The dissipation of this free energy by a civilization could then dominate the entropy production of the planetary system  in principle .  We could expect such a planetary state of  Earth in the future *if* *humanity successfully manages a transition to an energy system based entirely on solar energy*. Other forms of renewable energy with low feedback can also play a role.

Finally, even with highly pessimistic assumptions about the probabilities for the evolution of technological civilizations over the history of the Universe (Frank & Sullivan 2016), there likely have been many Class V planets across cosmic history (though we note that our galaxy could still be sterile now). Such dynamics might intentionally evolve by eco-evolutionary dynamics.

**IV. Eco-Evolutionary Dynamics and Planetary Transitions**

The hypothesis that planetary transitions from Class IV to V are plausible is supported by emerging evidence of eco-evolutionary feedbacks on contemporary time scales. Eco-evolutionary dynamics are reciprocal interactions of ecological and evolutionary processes over time scales shorter than evolutionary biologists used to assume were necessary (Pimentel 1961, Schoener 2011). It is well established that changes in ecological conditions may drive evolutionary change in species traits that then alter ecosystem function (Post &Palkovacs 2009). The reciprocal/simultaneous outcome of such interactions are only beginning to emerge (Matthews et al 2011). Furthermore, increasing evidence that humans drive major micro-evolutionary change implies human-driven phenotypic evolution might lead to ecosystem change on planetary scales (Palkovacs et al. 2012, Alberti, M. 2015). When such changes are successfully directed towards the establishment of long term (sustainable) versions of a planetary biosphere, what we term *"agency-dominated" biospheres*, then the planet enters Class V.

Research demonstrates how rapid evolution might affect ecosystem functions by changing functional traits—organisms' morphological, physiological, phenological, or behavioral characteristics that regulate their effects on ecosystems (Loreau 2010). Individual trait variation has significant implications for ecosystem productivity and stability. For example, the evolution of traits that regulate consumers' demand for resources affect nutrient cycling, and ultimately, the magnitude/spatial distribution of primary production (Matthews et al 2011). Evolution of traits of ecosystem-engineers (any organism that creates, significantly modifies, maintains or destroys a habitat), such as dune and marsh plants and mangroves, can affect their functional roles in maintaining the structures of estuarine and coastal environments.

With the emergence of the Anthropocene epoch, humans have become major selective agents capable of unleashing unprecedented evolutionary consequences (Alberti 2015). This is particularly evident in human-dominated environments (i.e emerging urban agglomerations, (Alberti 2016, Alberti et al 2017). Rapid urbanization affects eco-evolutionary dynamics both by changing habitat and biotic interactions and by accelerating transitions of economies toward increased demand for resources. Examples of eco-evolutionary feedbacks associated with urbanization have been documented for many species of birds, fish, plants, mammals, and invertebrates (Alberti 2015). Humans' selective pressures on traits alter the population dynamics of multiple prey species, reconfigure trophic interactions, and ultimately drive changes in community dynamics that control ecosystem functions (Matthews 2011).

The planetary impact of human activity— for instance measured by the free energy appropriated by humans to meet their metabolic activity (e.g., net primary production) and socioeconomic activity (primarily fossil fuel consumption)—is expected to increase in the future (Krausmann et al 2013). Yet, whether the increase in energy demand will be met by degrading the ability of the Earth system to generate free energy or enhance free energy generation within the Earth system, will depend on the capacity of humans to redirect current activities via technological innovations to increase efficiency of use of resources and solar radiation (Kleidon 2012).

Human-driven eco-evolutionary feedbacks provide novel opportunities for evolutionary innovation. These innovations could redefine interactions between our technologically-driven civilization and global ecological processes leading to unprecedented biospheric functions. From a thermodynamic perspective, these may involve novel free energy generation capacities. By extending the range of phenomena causing evolutionary change to include niche construction (Matthews 2011), scholars working at the interface of physics and evolutionary biology have hypothesized that evolution could expand the capacity of biological systems to dissipate free energy, thus maintaining and enhancing a state far from thermodynamic equilibrium (Loudon 2015). For example, Loudan et al. (England 2013) provided experimental evidence that evolution by niche construction affects dissipative ecosystem dynamics.

From an astrobiological perspective, all planetary transitions beyond Class II imply emergence of eco-evolutionary innovations. The emergence of oxygenic photosynthesis in Cyanobacteria represents one of the most remarkable evolutionary innovations in Earth's history. A transition to from a Class IV to a Class V world (through an Anthropocene will require humans or any technological species) to outperform microbes.  (Note that here we are explicitly using Anthropocene to refer to period when humans, or any technological species, begins dominating their planetary systems in a way that is unlikely to be sustainable).

One plausible scenario for this transition is planning of new biospheric functions. Alternatively, coupled human-natural systems self-organize to generate free energy as a result of eco-evolutionary processes. This implies a level of cooperation between agency (the technological species) and the biosphere that accounts for the inherent non-linearity and complexity of planetary systems.  Rethinking planetary evolution in the presence of humans implies expanding the notion of co-evolutionary causes.  This requires including ideas of niche construction and cultural co-evolution through both inheritance and social learning (Laland 2014). Understanding the role of coupled human-natural systems in planetary evolution would require bridging theories from geoscience, evolutionary biology and ecosystem science.

### V. Non-equilibrium Thermodynamic Measures of Planetary Transitions

Given our focus on non-equilibrium thermodynamics in coupled planetary systems, we seek measures of planetary-scale non-equilibrium or disequilibrium that capture the cascade of new functions and complexity occurring as planets make transitions (if they make them) upward in class[1].

The specification of planetary disequilibrium has a long history in astrobiology. The work of Lovelock and collaborators (Lovelock 1965; Lovelock 1975; Lovelock and Margulis 1974) led to recognition that the biosphere strongly influenced Earth's geochemical environment, including the composition of the atmosphere. The co-existence of long-term incompatible chemical species like oxygen and methane was proposed as one possible sign of life (Hitchcock and Lovelock 1967; Lederberg 1965; Lovelock 1965). Given that Earth's atmospheric gases are modulated by biology (Catling & Kasting 2007), it is reasonable to expect some planetary atmospheres to be similarly perturbed away from chemical equilibrium by biogenic gas fluxes. It is worth noting that chemical disequilibrium is often seen as the most promising means of identifying biospheres in exo-planet studies (Cockell *et al.* 2009). It has proven difficult, however, to quantify the signatures of biogenic non-equilibrium that could be detected from a distance (Simoncini *et al.* 2013; Seager *et al.* (2013), and to relate disequilibrium to the rate by which it is generated (Simoncini *et al*. 2013).

Recently, Krissansen-Totton et al. (2015) carried out detailed chemical modeling of all the planets in our solar system to determine their degrees of chemical disequilibrium. Lippincott *et al.* (1967) and Lovelock (1975) made early attempts to calculate such a thermodynamic disequilibrium for the Solar System planets. Their calculations were, however, hampered by incomplete data of atmospheric compositions and crude thermodynamic data. The final metric Krissansen-Totton et al. (2015) employed in their work was the Gibbs Free Energy ($\varphi$) normalized by the molar thermal energy $RT_p$ (where R is the gas constant and $T_p$ is the average planetary temperature for each world). Thus ($\varphi/RT_p$) provided a measure of chemical disequilibrium, normalized to the different levels of stellar flux received by planets at different orbital radii. Table 1 shows the Krissansen-Totton et al. (2015) values for two examples of a Class II planet (Venus and Mars) and Earth (as our example of a Class IV world). As shown, Earth has substantially higher values of ($\varphi$ /RT$_p$) than either of the Class II planets, as we would expect from discussion of the classification system.

---

[1] In most cases, however, planets will begin with primordial atmospheres due to outgassing. The timescale for atmospheric loss is determined by the planet's escape velocity and atmospheric composition. Thus, most planets begin as Class II and evolve down to Class I or continue upwards if biotic evolution begins.

To include planets in which a technological energy-harvesting species is active, however, we must consider a different metric. As shown in Kleidon (2011), it is possible to estimate the entire free energy or work budget for current Earth system processes. Beginning with a global solar irradiance of 1.6 x 10$^{17}$ W, Kleidon 2011 ( table 2) provides estimates of the magnitude of work done for various processes such as atmospheric circulation, hydrologic cycling and terrestrial biospheric productivity. Each form of work is categorized in terms of its source as *kinetic* (associated with velocities *v*), *potential* (associated with topography $\varphi$) or *chemical* (associated with chemical potentials $\mu$ or affinities *A*). For thepurpose of this research, we use chemical work as a measure of disequilibrium (Kliedon 2011, table 3). We make this choice because of the relationship between novel forms of free energy generation within a planetary system and the capacity of evolution to generate novel functions that then feed back to ecosystems (section IV). For example, the evolution of photosynthesis allowed direct chemical capture (and use) of energy locked in solar photons in the early Earth Systems. This was more efficient than heat-engine forms of capture and use represented by mechanical mean to generate disequilibrium. As biotic evolution progressed, the feedback and coupling to ecosystems, (both locally and cascading up to the biosphere as a whole), allowed for innovation which carried Earth from a thin biosphere (Class III) to its present hybrid state and, hopefully towards a world that can sustain the presence of a long-lived energy-intensive technological civilization.

With emphasis on generation rate of chemical free energy and work, we have chosen to use values for the Earth at different points in its evolution as a standard. Thus, for a Class II planet without a biosphere, we use only abiotic components of chemical work on *the present-day Earth*. This is likely an overestimate since the biosphere modifies the factors driving chemical free energy. It allows a comparison, however, between different modes of free energy generation, as shown below.

For our characterization of a Class III world as those possessing a "thin" biosphere, we use the primary productivity of Earth during the Archean eon, before the establishment of an oxygen rich atmosphere. Canfield (2005) calculated the total productivity from anoxygenic photosynthesis during this period as approximately 5% of the Earth's current primary productivity. We use this estimate to scale the values of net productivity in both terrestrial and oceanic systems given in Kleidon (2011).

Considering Earth at the start of the proposed Anthropocene epoch to be in a hybrid state transitioning between a Class IV and Class V planet, we consider the work/free energy utilized by humans in the production of chemical disequilibrium. Thus, we use the total human appropriation of primary productivity and our primary energy consumption given in Kleidon (2011) as measures of "technological" chemical disequilibrium generation.

Finally, we consider what might occur on a Class V planet in which energy is harvested and utilize,d but in a way imposing the smallest climatic impact possible. As discussed in Section III, this may entail development of novel ecosystem functions occurring through the action of a technological species. The species establishes an "agency-dominated" biosphere that provides support for the civilization while maintaining its own viability. Given the enormous uncertainties associated with this process, we consider two scenarios associated with alteration of the Sahara region. Given the Sahara's large area (A = 9 x $10^{12}$ m$^2$) and current desert state, it may serve as future site for developing large scale, (i.e planetary) projects aimed at sustainable energy harvesting or biospheric adaptation. Before we begin, we note the total power used by human civilization which a conservative estimate gives as $P_{civ}$ = 22 x $10^{12}$W (Kleidon 2016).

We first consider the case of simple photovoltaic coverage of the Sahara. Using current industrial grade efficiency of photovoltaic panels (~ 20%) and a mean solar radiation of about 200 W m$^{-2}$ , we find an electric energy (i.e free energy) generation of P$_{civ}$ = 360 x 10$^{12}$ W ( this is about 16 times the current worldwide primary energy consumption).

One could also consider "greening" the Sahara.  The large-scale cultivation of desert regions for both the production biomass energy and to produce beneficial climate feedbacks has been studied by a number of authors (e.g., Ornstein et al. 2009; Becker et al. 2013; Bowring et al. 2014).  It thus appears as a crude version of a candidate process indicative of a Class V planet as it requires the supply of water by technological means (e.g. desalination of seawater).  To calculate the free energy generation from greening the Sahara, we follow Bowring et al. (2014).  Using a maximum photosynthetic efficiency of 3 % and an efficiency of 1.5% for biomass production (to account for losses by autotrophic respiration) we find a chemical free energy generation (in form of biomass) of P$_{chem}$ = 27 x 10$^{12}$W.  Using the mean energy content of sugar, this corresponds to about 10.7 x 10$^{15}$ gC/yr, which is about 18% of the current productivity on land.  To accomplish the cultivation of the desert would require about 1.1 x 10$^{13}$ m$^3$/yr of water (Bowring et al. 2014, this corresponds to 15% of the natural evaporative flux on land).  Since sea water would likely be required, we must include the energetic cost of desalination which requires a minimum of 3.8 MJ/m$^3$ (a more realistic value of 14.5 MJ/m$^3$, see Elimelech and Phillip, 2011).  The energetic expense for desalination is therefore 1.3 - 5 x 10$^{12}$ W.  Thus, we find that the energetic gain of about 27 TW outweighs the energetic cost of 1.3 - 5 TW by a factor of 5.4 - 20.8.  In other words, the Energy Return On Investment, EROI = 5.4 - 20.8.  Using the more conservative estimate of P$_{civ}$ = 22 TW, we find that greening the Sahara yields more free energy generation than what is currently consumed by humans in terms of primary energy and the human appropriation of net primary productivity.  Thus the total would be 22 + 27 TW or 49 TW.

Table 1 and Figure 5 show  results where the chemical free energy budgets are shown for each planetary class.  Class I worlds with no atmosphere produce no chemical free energy.  Class II worlds have only fluid systems (atmosphere and liquids) at their disposal but can still generate chemical free energy (and hence disequilibrium) via processes like weathering (i.e. runoff) or processes within liquid systems (i.e. desalination). A Class III world also has these processes at work and so we carry forward the value of abiotic free energy generation used for the Class II world. The Class III world, however, also includes biotic activity. Using the net productivity of the Achaean Earth (Canfield 2005), we see that Class III worlds have a total chemical free energy budget higher than that for a Class II world. Note that for this class world, the contribution from the biotic component is smaller than that from abiotic sources. Moving forward, unlike the thin biosphere of a Class III world, the biotic activity in a "thick" system in a Class IV world (as in Earth before the advent of the Anthropocene as defined in the introduction) represents a significant fraction of the planet's total chemical free energy budget. In a Class V world, large-scale agent based biosphere adaptation via, for example, "desert greening" can increase the free energy available to a civilization by significant factor, while the capture of even a small fraction of stellar incident photons yields the highest free energy budget of all. Finally, a hybrid planet like the Earth as it deepens it's entry into the Anthropocene shows technological activity already capturing and contributing a significant fraction of the total chemical energy budget for the planet.

**VI. Conclusions**

The situation humanity finds itself at entry to the Anthropocene is likely not unique when seen it its proper Astrobiological context. As long as the probability for energy intensive civilization evolution is $> 10^{-22}$ per habitable zone planet, then humanity is not the only example of such evolution. Since the laws of thermodynamics hold for all planets, the kinds of feedback associated with the Anthropocene must also have occurred elsewhere. While the response of any given civilization will vary in ways that will depend on social, cultural, and other factors well beyond prediction, critical aspects of a successful response lie purely in the domain of planetary scale thermodynamics through a combination of physics and chemistry. Thus, a classification of planet types that includes sustainable biospheres shaped by the agency of technological civilization is both meaningful and useful. Our classification is based on thermodynamic considerations of different forms of free energy generation and disequilibrium states, clarifying life's role in altering planetary systems and yielding greater levels of disequilibrium.

In all cases, the transition from one planetary class to another involves some form of evolutionary innovation. This innovation is made possible via coupled interactions between evolution and ecosystems resulting from the synchronization of timescales for the emergence of new genotype/phenotypes and novel ecosystem functions. We argue that the addition of new processes (biospheric functions) to the coupled planetary systems leaves imprints in the behavior (and hence dynamic states) of those systems. For the transition I->II, the innovation is simply the additional degrees of freedom made possible by the addition of fluid components to the coupled systems in the form of an atmosphere and, potentially, a hydrosphere. In the other transitions, the innovations are truly evolutionary as novel functions/behaviors emerge from within the coupled systems. For the transition II-III, innovation involved the evolution of an entirely new system (the biosphere) while in transition III-IV and IV-V the innovation was driven by evolution within the biosphere. It might be argued that the IV-V transition involves the addition of a "technosphere" or "noosphere" which might eventually grow in extent and autonomy that it should also be considered as a separate system.

The classification system outlined in this paper allows Earth's entry into the proposed? Anthropocene to be seen as a hybridization in the transition between different types of planets from IV to V. We believe this astrobiological perspective is an essential expansion of the discussions of the Anthropocene and paths towards planetary sustainability, because any world hosting a long-lived energy-intensive civilization must share at least some similarities in terms of the thermodynamic properties of the planetary system. Understanding these properties, even in the broadest outlines, can help us understand which direction we must aim our efforts in developing a sustainable human civilization.

Table 1

|  | Abiotic | Biotic | Technological | $Z = \varphi/RT\_p$ | Notes |
|---|---|---|---|---|---|
| Planetary Type |  |  |  |  |  |
| I | 0 | 0 | 0 |  |  |
| II | 2.8E+01[1] | 0 | 0 | ~0.08 Mars<br>~0.008 Venus |  |
| III | 2.8E+01 | 1.1E+00[2] | 0 |  | Class III biotic = 5% current Earth value |
| IV | 2.8E+01 | 2.5E+02[3] | 0 | ~3.0 Earth |  |
| Hybrid | 2.8E+01 | 2.5E+02 | 2.2E+01[4] |  |  |
| V | 2.8E+01 | 2.5E+02 | 4.9 E+01 (DG + current value)<br><br>3.8E+02 (PC+current value) |  | Class V Technological estimates should be considered lower limits |

Table 1.0 Chemical Free Energy (Work) Budget in TW vs. Planetary Type. We show two estimates of free energy for Class V planets based on modifications of the Sahara. DG corresponds to desert greening while PC refers to photovoltaic coverage. Table also includes dimensionless disequilibrium (Z) for Mars and Earth from Krissansen-Totton et al 2015.

[1]From Kleidon 2011 Table 3. This value equals sum of desalinization (27 TW) and runoff (~1 TW).
[2]From Canfield (2005) Table 1 where I estimated that he was finding a 5% of current Earth primary productivity (see next footnote)
[3]From Kleidon 2011 Table 3. This value is sum of terrestrial (152 TW) and marine (63 TW) productivity.
[4]From Kleidon 2016

Figure captions

Figure 1. Schematic showing generation of different forms of free energy (boxes in the center) via absorption of low entropy stellar radiation. Differential radiative heating/cooling provide gradients for planetary heat engines ($P_{clim}$), generating motion. Photochemistry/geochemistry generate chemical free energy at the rate $P_{chem}$, Photosynthesis generates chemical free energy at a rate $P_{bio}$ to fuel life. Last, a technological civilization relies on a rate of free energy appropriation $P_{civ}$ from the biosphere and by technological means (i.e. photovoltaics). These processes feed back to the radiative forcing of the planet by heat transport and modified radiative characteristics associated with atmospheric composition (e.g., $O_3$, $CO_2$, and $CH_4$), or with modified surface properties (i.e. vegetation).

Figure 2. Dominant processes for free energy generation and energy dissipation for different class planets. Dissipation increases with class as new processes are added to the planetary systems (see text for details). For Class V, an agency-dominated biosphere holds the planetary-systems within acceptable boundaries for energy intensive technological civilization.

Figure 3. Classes of planets with different abilities to generate free energy by different forms (see text for details)

Figure 4. Free Energy Generation for Chemical Disequilibrium vs Planet Class Chemical free energy from different sources across the five planetary classes. Yellow corresponds to purely abiotic processes. Brown corresponds to purely biotic processes. Blue corresponds to technological processes. Note that all abiotic values are taken current values for Earth. Class III biotic chemical free energy taken from estimates from Archean Earth net primary productivity (Canfield 2005); Class IV biotic chemical free energy taken from current values for Earth (Kleidon 2010); Hybrid Planet (HP) technologically sourced chemical free energy taken as net current human consumption. Class V technologically sourced chemical free taken as covering the Sahara with photovoltaics.

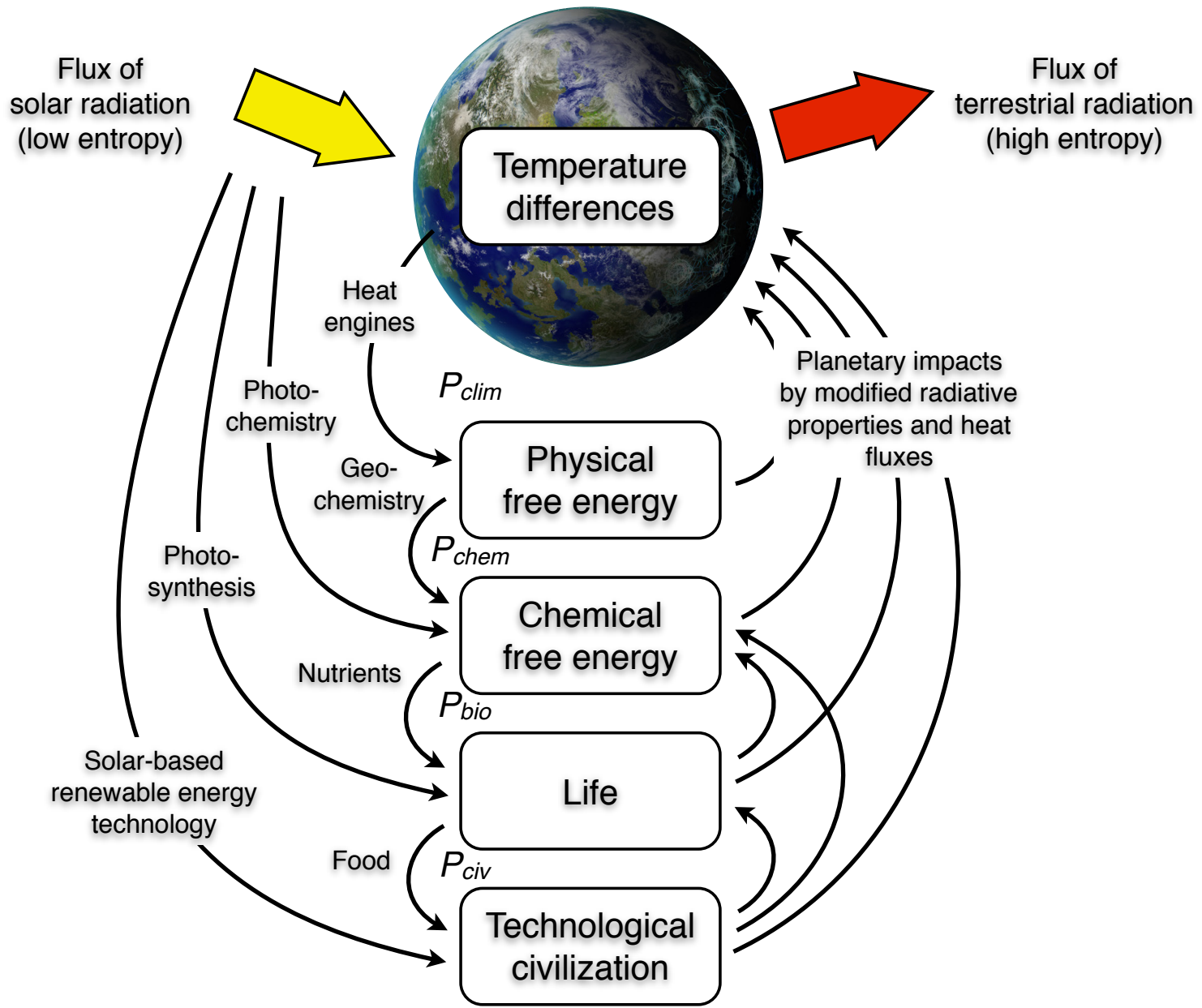

Figure 1

Figure 2

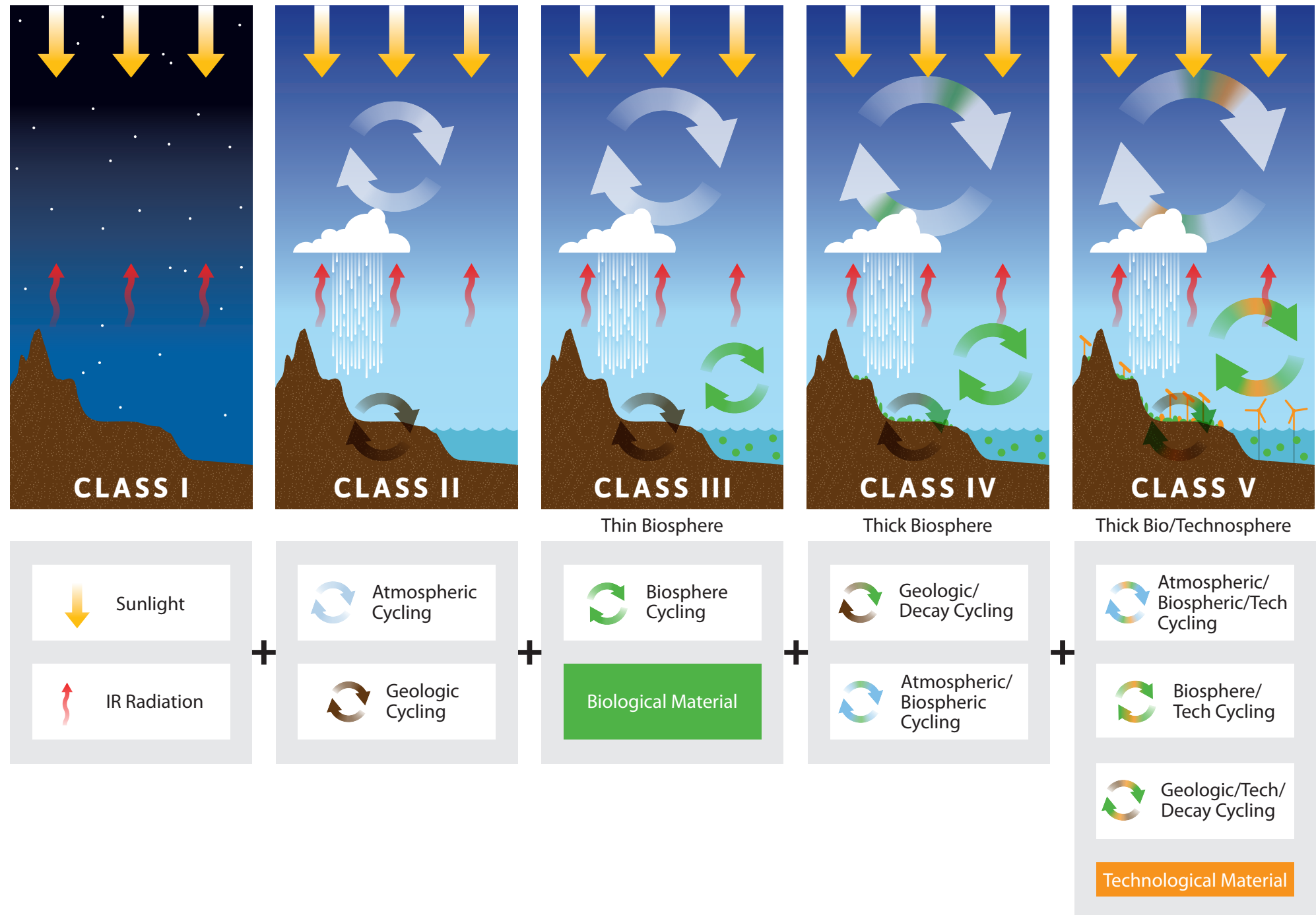

Figure 3

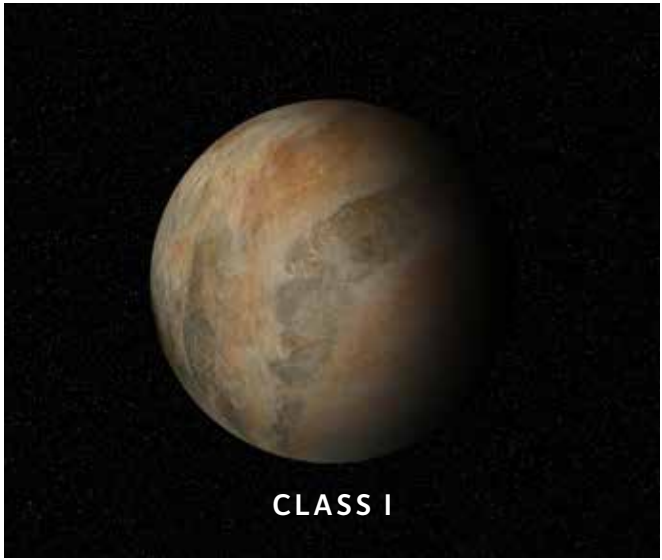

CLASS I

$P_{RAD} \gg 0$

$P_{CLIM} = P_{CHEM} = P_{BIO} = P_{CIV} = 0$

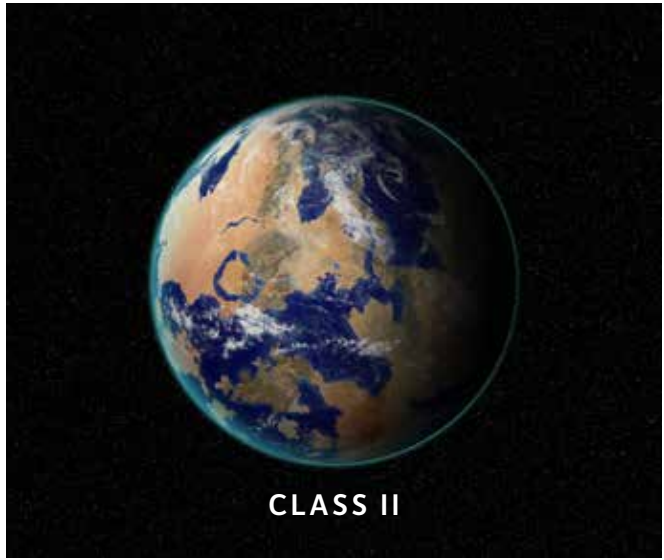

CLASS II

$P_{RAD} > 0$

$P_{CLIM} \gg 0, P_{CHEM} \geq 0; P_{BIO} = P_{CIV} = 0$

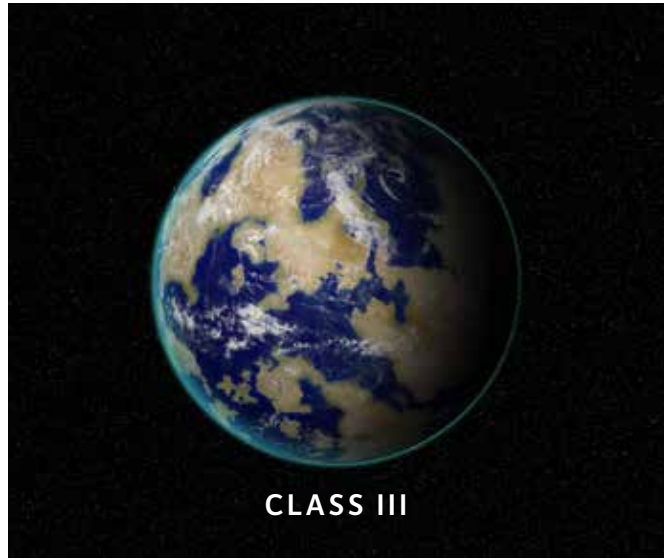

CLASS III

$P_{RAD} \gg 0$

$P_{CLIM} \gg 0, P_{CHEM} > 0, P_{BIO} > 0, P_{CIV} = 0$

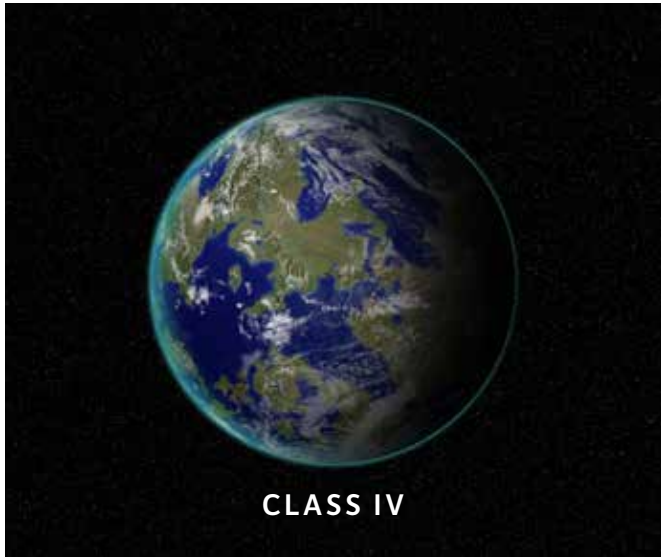

CLASS IV

$P_{RAD} \gg 0$

$P_{CLIM} \gg 0, P_{CHEM} \gg 0, P_{BIO} \gg 0, P_{CIV} = 0$

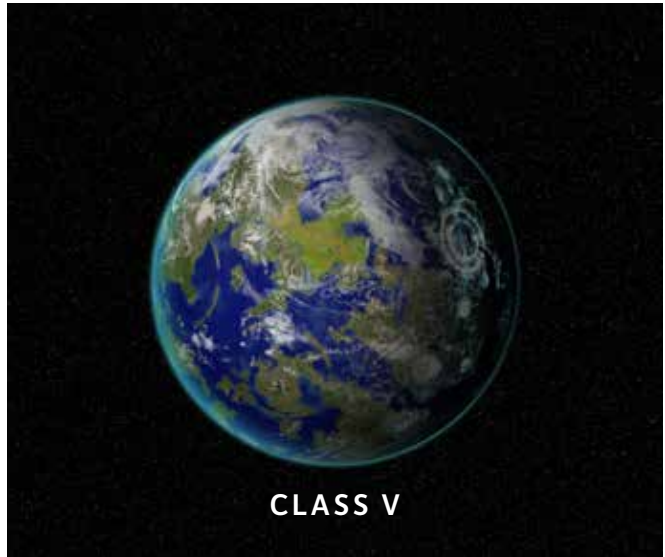

CLASS V

$P_{RAD} \gg 0$

$P_{CLIM} \gg 0, P_{CHEM} \gg 0, P_{BIO} \gg 0, P_{CIV} \gg 0$

Figure 4. Free Energy Generation for Chemical Disequilibrium vs Planet Class

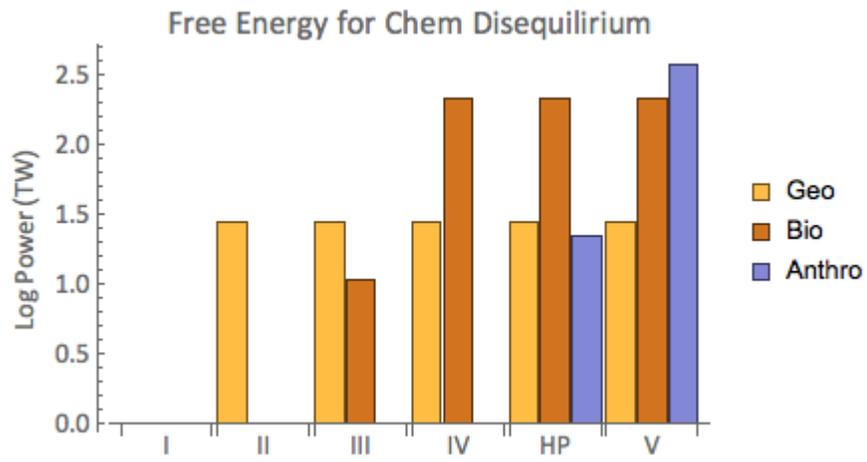